\documentclass{icrc29}
\usepackage{graphicx,amssymb,amsmath,times}

\begin{document}

\title[HiRes-II Systematics]{Studies of Systematic Uncertainties in the HiRes-II
\\       Measurement of the UHECR Spectrum.}
\author[A. Zech for HiRes]{A. Zech, for the HiRes Collaboration\\
              LPNHE, Universit\'{e} de Paris VI, 4 place Jussieu, F 75252 Paris Cedex 05, France    
}
\presenter{Presenter: A. Zech (aszech@lpnhep.in2p3.fr),\
fra-zech-A-abs1-he14-poster}

\maketitle

\begin{abstract}

We present studies of systematic uncertainties in the 
measurement of the ultra-high energy cosmic ray (UHECR) spectrum with 
the FADC detector of the High Resolution Fly's Eye experiment
(HiRes-II). One source of uncertainties lies in the simulation 
of the energy dependent aperture of the air fluorescence 
detector. We study the impact of changes in the energy spectrum
and composition that are used as input to the aperture simulation.
We also compare aperture estimates for two different hadronic
interaction models - QGSJet and SIBYLL. Systematic uncertainties may
further be introduced by the modeling of the aerosol component of the atmosphere. 
We have repeated the HiRes-II monocular analysis using an atmospheric database 
with hourly entries instead of our measurement of the average aerosol 
content. We will discuss changes in reconstructed energies and in the 
resulting spectrum.

\end{abstract}

\section{Introduction}

The two HiRes detectors measure the flux of cosmic rays from $\sim 10^{17}$ eV up to the highest energies by observing the cascades of secondary charged particles generated in the atmosphere. 
Nitrogen molecules in the path of the particle ``shower'' are excited and emit fluorescence photons
in the UV range, which can be detected with photomultiplier tubes.

The main systematic uncertainties that are introduced in the measurement of the ultra-high energy
cosmic ray spectrum with the HiRes experiment have been reported in \cite{HRMono-PRL}.
Including uncertainties in the phototube calibration, fluorescence yield, ``missing
energy'' correction and aerosol concentration in the atmosphere, the
total systematic uncertainty in the measured flux is $31 \%$ for each of the two 
monocular spectrum measurements.

For this paper, we have used the simulation and reconstruction programs of the HiRes-II monocular analysis to  examine additional systematics that can affect the simulation of the HiRes aperture and the reconstructed energies. Since the aperture of an air fluorescence detector is a function of the energy of 
the observed cosmic rays, it has to be modeled very carefully with the help of detailed Monte Carlo (MC)
simulations. We have considered the effects of varying the different assumptions that are adopted in our
simulation programs. We have studied variations in the input energy spectrum and composition,
the hadronic interaction models, and the description of the aerosol content of the atmosphere.

\section{Input Energy Spectrum}
The energy distribution measured by the detector is a convolution of the energy distribution of cosmic rays at their arrival on Earth with the detector response, i.e. the efficiency of the detector and its finite resolution. In the process of unfolding \cite{cowan} the measured energy spectrum, we use realistic MC simulations of both the air shower development and 
the complete measurement process -- including light generation and propagation, the optical, electronic
and trigger system of the detector -- to determine the aperture of the experiment as a function of the
cosmic ray energy. It is important that the energy spectrum that is assumed as an input to the simulation programs is matched in shape to the measured spectrum. By doing so, one avoids a bias stemming 
from the limited detector resolution. 
We have accounted for effects of finite energy resolution on the determination of the aperture by using an assumed input spectrum that resembles our measured spectrum. 

\section{Input Composition}
In our standard analysis, we have generated libraries of air shower profiles, using the CORSIKA \cite{CORSIKA} and QGSJet \cite{QGSJet} programs, at different energies, and for proton and iron primaries.
The fractions of air showers initiated by proton and iron cosmic rays in our MC are determined from composition measurements by the {\it HiRes/MIA} and {\it HiRes} stereo experiments \cite{HRComposition}. In a given energy bin, we calculate the fraction of proton showers by comparing the mean of the depths of the shower maxima ($X_{max}$) recorded in the data to the mean $X_{max}$ predicted by 
the proton and iron showers in our library. The proton fraction depends therefore on the model used to simulate the library showers. However, the mean $X_{max}$ of the simulated events is given by the  {\it HiRes/MIA} and {\it HiRes} stereo data -- independently of the models in use.

\begin{figure}[htb]
\begin{minipage}[t]{0.48\textwidth}
\mbox{}\\
\centerline{\includegraphics[width=\textwidth]{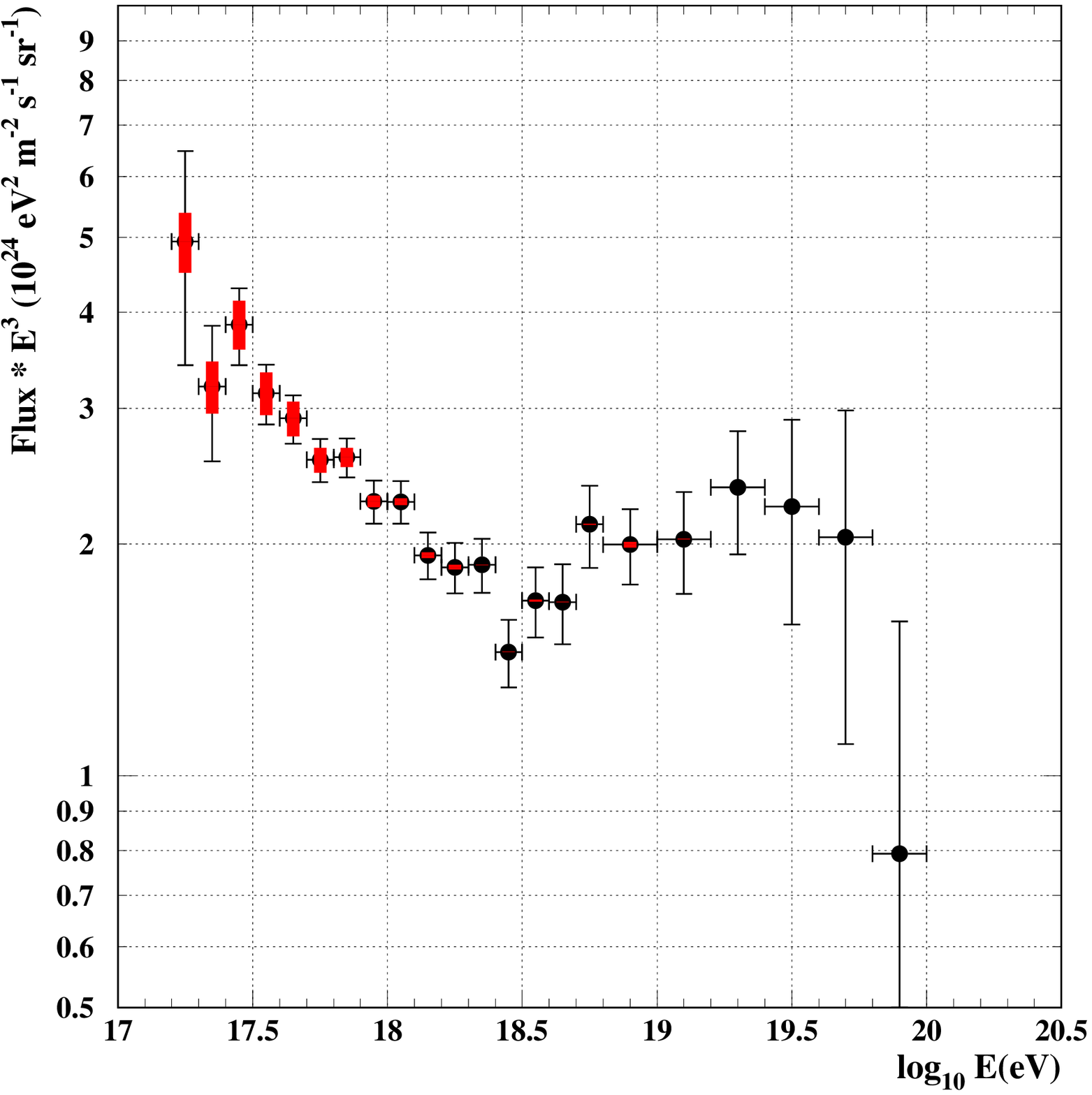}}
 \caption[HiRes-II spectrum with composition error bars.]
 {HiRes-II energy spectrum with systematic uncertainties
    (thick error bars) corresponding to a $\pm 5 \%$ change in the
    proton fraction of the MC.}
\label{compsys}    
\end{minipage}
\hfill
\begin{minipage}[t]{0.48\textwidth}
\mbox{}\\
\centerline{\includegraphics[width=\textwidth]{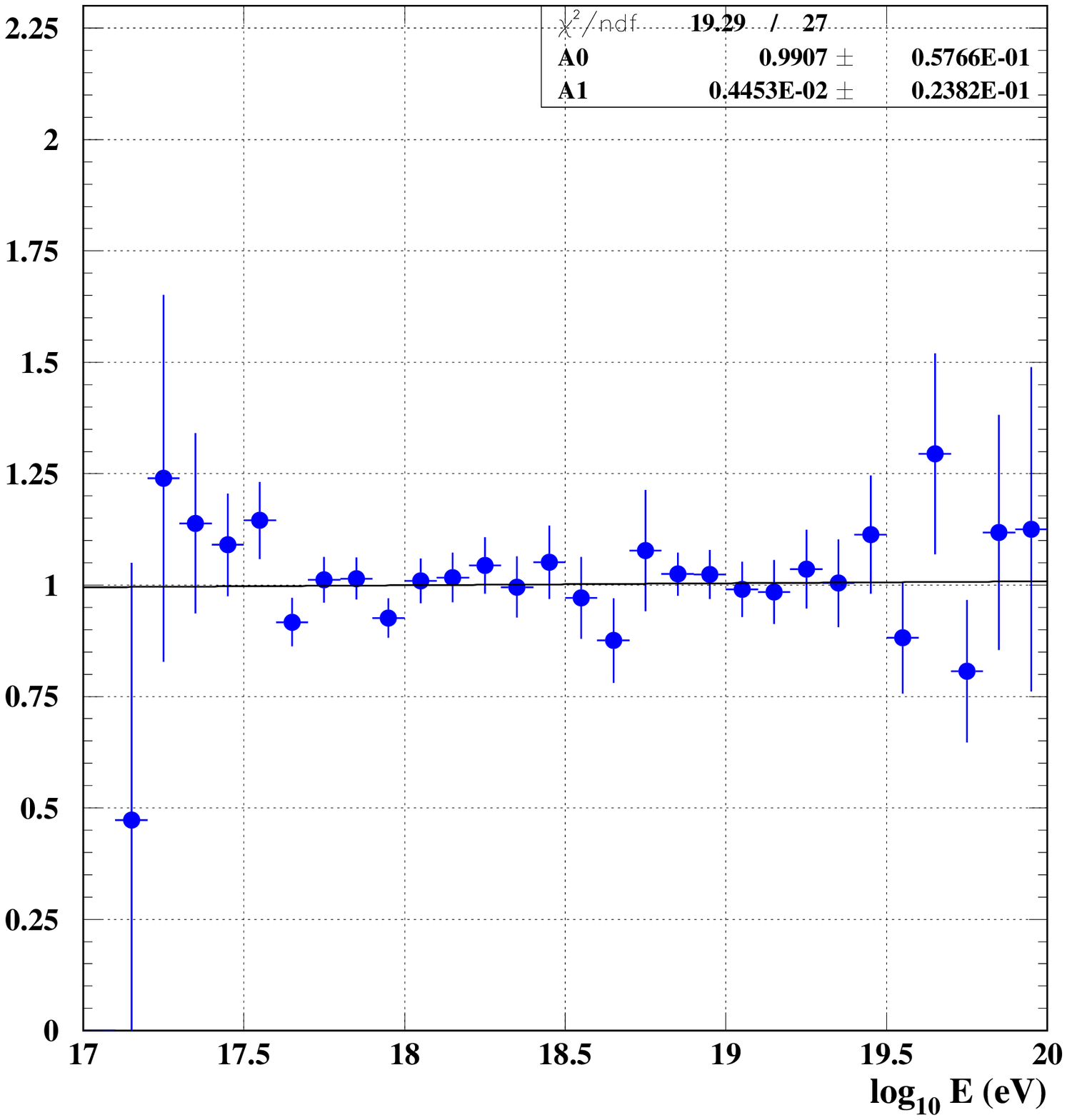}}
  \caption[Ratio of SIBYLL and QGSJet acceptances.]
{Ratio of the acceptances estimated using SIBYLL (numerator) and QGSJet (denominator).
A line has been fitted to the ratio.}
\label{hadratio}
\end{minipage}
\end{figure}

Due to the limited coverage in elevation angle of the HiRes detectors ($3^{\circ} - 31^{\circ}$ for HiRes-II) , changing the composition of the cosmic ray flux in the simulation without changing the model of the shower generator can affect the simulated aperture. At the same energy, iron showers develop higher up in the atmosphere than proton showers (i.e. they have a smaller $X_{max}$) , and thus are more likely to be above the elevation coverage of the detectors. 
Using the QGSJet model, we have calculated the effect that a change of the assumed proton fraction would have on the simulated aperture. A change in the proton fraction
of $\pm 5$ \%, which is the magnitude of the relevant uncertainties 
in the {\it HiRes/MIA} measurement, leads to systematic uncertainties smaller than the statistical uncertainties in the HiRes-II spectrum \cite{CRIS-AZ}. At energies above $\sim 10^{18}$ eV, where observed showers are
on the average farther away from the detector, the effect becomes negligible (see Figure\ref{compsys}).

\section{Hadronic Interaction Models}
The main uncertainty in the air shower simulations comes from our limited knowledge 
of the initial hadronic interactions, which take place 
at energies by far exceeding those that can be observed in the laboratory. In order to estimate 
the influence of those uncertainties on the simulation of the aperture, we have generated two
libraries of air showers within the CORSIKA framework, using two different hadronic interaction models, QGSJet 1.0 and SIBYLL 2.1 \cite{SIBYLL}.  Showers simulated with SIBYLL have on the average larger $X_{max}$ values and slightly different elongation rates. Therefore, we had to re-calculate the
proton fraction that corresponds to the {\it HiRes/MIA} and {\it HiRes} stereo data points, and adjust the input composition to contain a larger fraction of iron showers in the case of the simulation with SIBYLL. By adjusting the proton fraction for the different models, we guarantee that showers are
generated at the same atmospheric depth, given by measurements.

The predictions for the ``missing energy'', i.e. energy that does not contribute to the ionization of air molecules and remains invisible for HiRes, show another difference: SIBYLL predicts a roughly 2 \% smaller ``missing energy'' fraction compared to QGSJet. We have taken this difference into account in the energy reconstruction of triggered MC events

We did not find any significant differences in our comparisons of data and MC events with the two MC sets. We have used the two MC sets to calculate a ratio of apertures, which gives us an estimate of the model dependent systematic uncertainties in our monocular spectrum measurement. As can be
seen from Figure \ref{hadratio},  there is no significant difference between the estimated apertures.
   
\begin{figure}[htb]
\begin{minipage}[t]{0.48\textwidth}
\mbox{}\\
\centerline{\includegraphics[width=\textwidth]{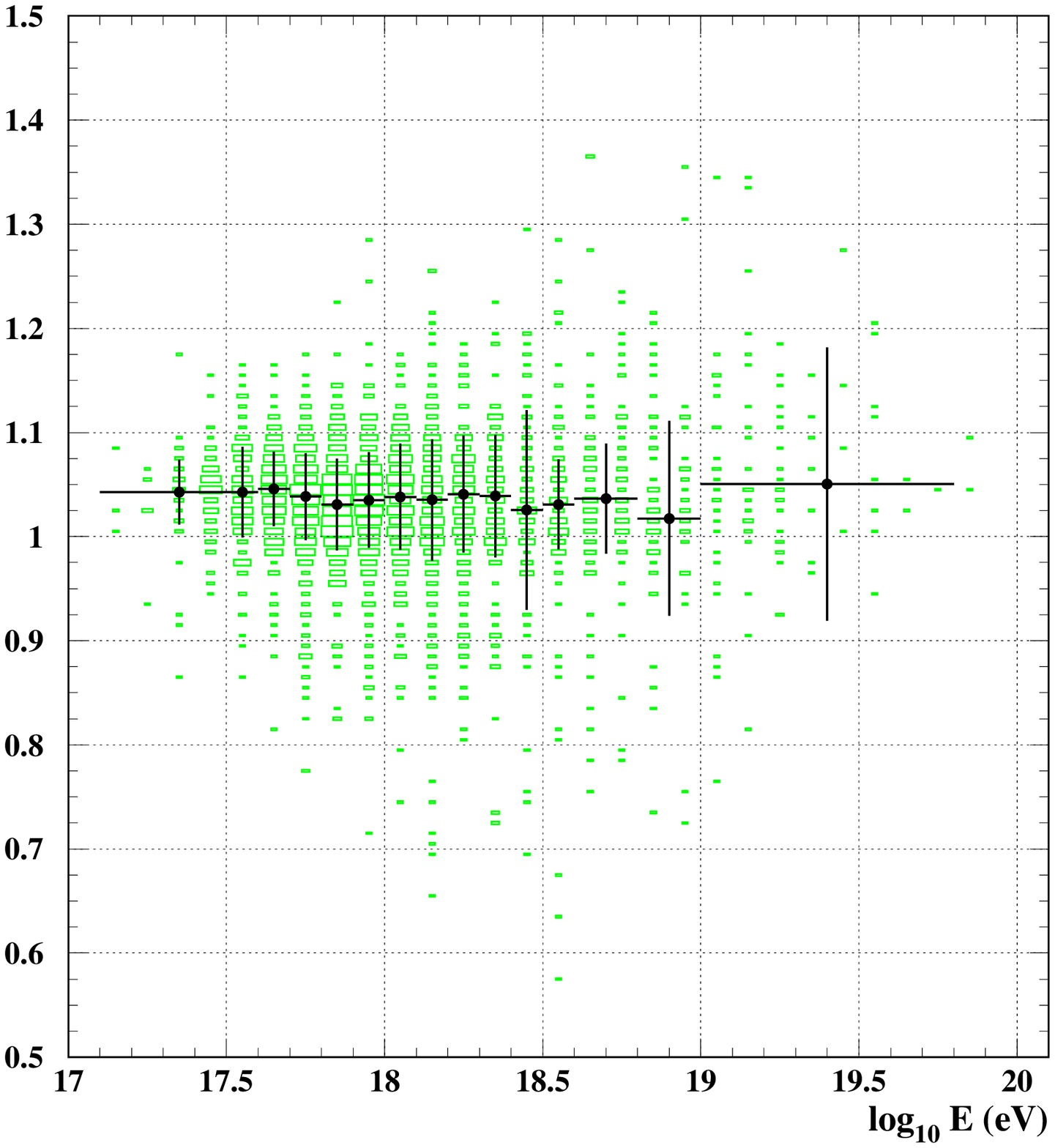}}
  \caption{Ratio of event energies reconstructed with an average atmosphere (numerator)
and with the atmospheric database (denominator) vs. event energy (with average atmosphere). 
The scatter of points is shown as well.}
\label{atmosratio}
\end{minipage}
\hfill
\begin{minipage}[t]{0.48\textwidth}
\mbox{}\\
\centerline{\includegraphics[width=\textwidth]{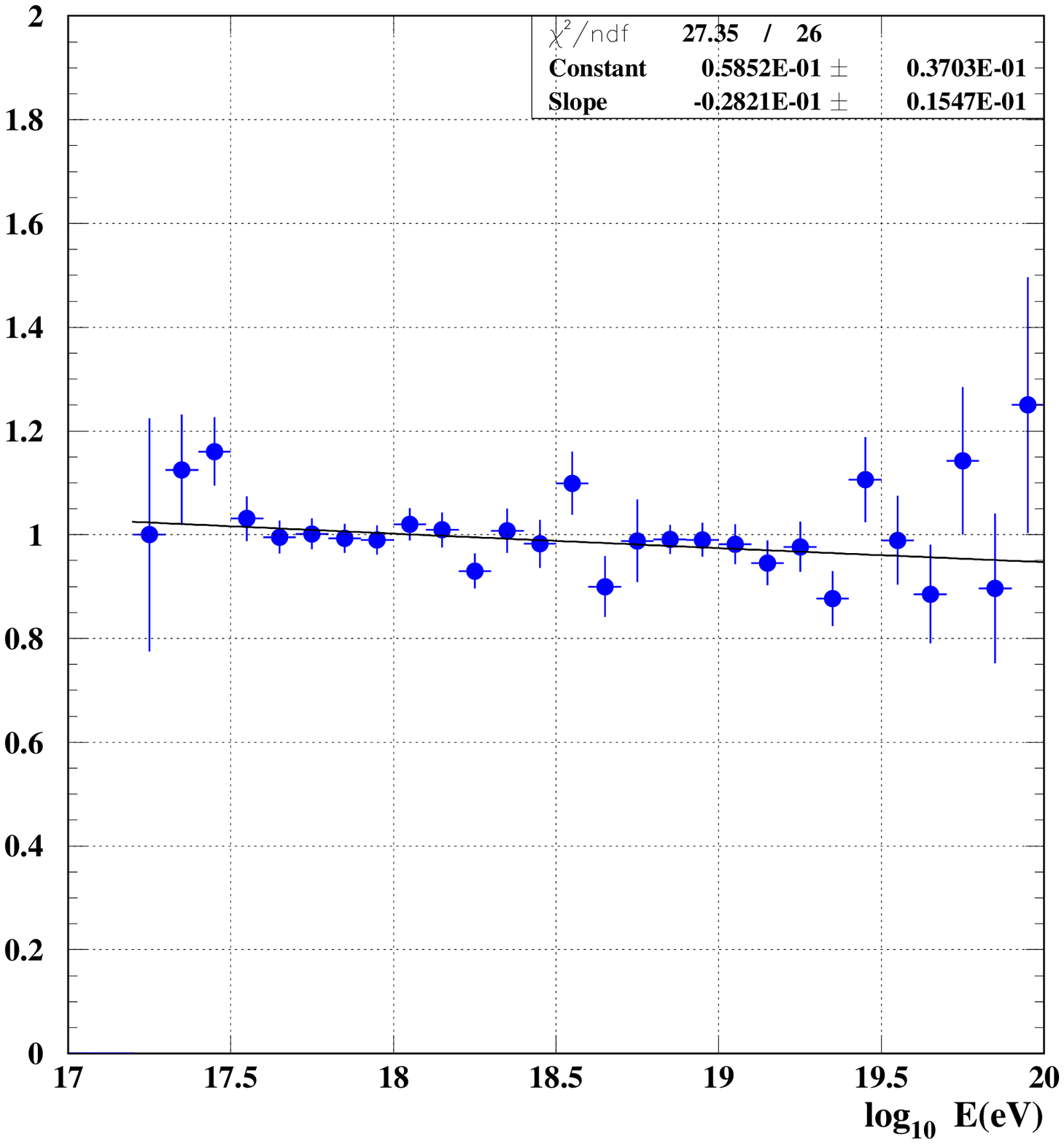}}
  \caption{Ratio of acceptances for a MC set generated using an
           atmospheric database and reconstructed with the
           database (numerator) and an average atmosphere (denominator).}
  \label{apratio_adb}
\end{minipage}
\end{figure} 
    
\section{Atmospheric Database}
For the monocular energy spectra published in \cite{HRMono-PRL}, a measurement of the average vertical aerosol optical depth (VAOD) and attenuation length was used in our analysis to 
define the aerosol density of the atmosphere during the period of data acquisition. 
We have repeated the analysis of HiRes-II data using a database with hourly entries of the 
measured VAOD and attenuation length, instead of the average values. 

Figure \ref{atmosratio} shows the changes in the reconstructed energies that are introduced with
the atmospheric database for data recorded by HiRes-II in 17 months between December 1999 and September
2001. Gaussian fits have been applied to the difference between the reconstructed values with
and without database. The points represent the Gaussian means for different energy bins; the 
error bars are standard deviations. The energies reconstructed with database are on the 
average $\sim 4 \%$ smaller. This is due to a slightly clearer atmosphere determined with an
improved analysis method for the VAOD values that went into the database (see \cite{atmospaper}).
To calculate systematic effects on the spectrum, we have also simulated an aperture
with use of the atmospheric database and compared it to the estimate using the average value.
We have not found any significant differences in the  apertures, as can be seen in Figure \ref{apratio_adb} \cite{CRIS-AZ}.

\section{Conclusions}
None of the sources of possible systematic uncertainties we have studied here contribute 
significantly to our 
published estimate. We have seen that adjusting the input energy spectrum to the shape of the measured spectrum is an important step in the deconvolution of the cosmic ray spectrum. 
Our simulated aperture is sensitive to the assumed input composition for energies below $\sim 
10^{18}$ eV. By using a measured composition as an input to our simulation programs, 
our analysis does not depend strongly on the assumed hadronic interaction model.
For the 17 month period tested here, the description of the aerosol density using an hourly
database does not cause any significant differences in the spectrum, when compared with 
an average atmosphere. 

\section{Acknowledgements}
This work is supported by US NSF grants PHY-9321949, 
PHY-9322298, PHY-9904048, PHY-9974537, PHY-0098826, 
PHY-0140688, PHY-0245428, PHY-0305516, PHY-0307098,  
and by the DOE grant FG03-92ER40732. We gratefully 
acknowledge the contributions from the technical 
staffs of our home institutions. The cooperation of 
Colonels E.~Fischer and G.~Harter, the US Army, and 
the Dugway Proving Ground staff is greatly appreciated.

\end{document}